\documentstyle[aps,prb,multicol,epsf]{revtex}   
\newcommand{\beq}{\begin{equation}}
\newcommand{\eeq}{\end{equation}}
\newcommand{\beqa}{\begin{eqnarray}}
\newcommand{\eeqa}{\end{eqnarray}}
\newcommand{\w}{\omega}

\renewcommand{\Im}{\mbox{Im} }

\begin{document}           

\title{Interaction corrections at intermediate temperatures: 
magneto-resistance in parallel field} 
\author{G\'abor Zala$^{(a,b)}$, B.N. Narozhny$^{(a),*}$, 
and I.L. Aleiner$^{(a,b)}$ }
\address{
$^{(a)}$Department of Physics and Astronomy, SUNY at Stony Brook, 
Stony Brook, NY, 11794, USA \\
$^{(b)}$Physics Department, Lancaster University, LA1 4YB, Lancaster, UK 
\\
} 
\date{\today} 
\maketitle 
\begin{abstract}                
We consider the correction to conductivity of a 2D electron gas due to 
electron-electron interaction in the parallel magnetic field at 
arbitrary relation between temperature and the elastic mean free time. 
The correction
exhibits non-trivial dependence on both temperature and the field. This
dependence is determined by the Fermi liquid constant, which accounts for
the spin-exchange interaction. In particular, the sign of the slope of the
temperature dependence is not universal and can change with the increase 
of the field.
\end{abstract} 
\draft 

\begin{multicols}{2}  

{\em Introduction} -- In a 
previous publication \cite{us1} we have developed a theoretical
framework for studying interaction corrections to conductivity of the
two dimensional electron gas (2DEG) due to electron-electron
interactions for the arbitrary relation between temperature $T$ and 
the elastic mean free time $\tau$. 
To describe strong coupling between electrons we used
the conventional Fermi liquid \cite{lfl} constants. In particular, we
found \cite{us1} that the temperature dependence of the longitudinal
conductivity in a 2DEG is determined by a single Fermi liquid constant
$F_0^\sigma$, which describes the strength of the spin-exchange
interaction. In principle, its value can be found from a measurement
of the Pauli spin susceptibility
  
\begin{equation} 
\chi = \frac{g^2\mu_B^2\nu}{1+F_0^\sigma},  
\label{pauli} 
\end{equation} 

\noindent  
where $\mu_B$ is the Bohr magneton, $g$ is the bare electron Lande factor
and the density of states $\nu$ should be obtained from a
measurement of the specific heat (at $\tau^{-1}\ll T \ll E_F$).
Unfortunately, to the best of our knowledge no measurement of the spin
susceptibility has been reported for the 2DEG created at the interface
of a semiconductor heterostructure, which currently is the most common
type of an experimental sample \cite{ex0}. However, we have conjectured
\cite{us1} that the same constant $F_0^\sigma$ should describe the
transport properties of the 2DEG in an external magnetic field. In
this paper we address the case of the parallel magnetic field and
calculate the magneto-conductivity. The case of the perpendicular
magnetic field and the theory of the Hall resistance was
discussed in a recent publication \cite{hall}.
 
Early
theoretical efforts focused on calculating the magneto-conductivity
within the diffusive approximation \cite{aar,lrr,lrp,mil}. While perfectly
justified for metallic thin films, this approximation might be
inappropriate for understanding of recent experiments\cite{ex0} in
semiconductor heterostructures, since these measurements are performed
in a regime where the temperature $T$ is of the same order of
magnitude as the inverse scattering time $\tau^{-1}$ (obtained from
the Drude conductivity). The opposite, ballistic limit was considered
recently in Refs.~\onlinecite{gdp,her}. While giving a reasonable description
of the magneto-resistance for weak interaction and
at small fields, the authors of Refs.~\onlinecite{gdp,her} did not realize
that both the temperature and the magnetic field dependence arise due to
large-distance (as compared with $\lambda_F$) processes and therefore
did not account for Fermi liquid renormalizations. The resulting 
temperature dependence of the conductivity (also see Ref.~\onlinecite{gdo})
is qualitatively erroneous: in particular
for the case of the fully polarized system Ref.~\onlinecite{gdp} has 
the incorrect sign and Ref.~\onlinecite{her} finds no temperature dependence
at all.
(see Ref.~\onlinecite{us1} for more details). In this paper we calculate the
magneto-conductivity for an arbitrary relation between $T$, $\tau$,
and the Zeeman energy $E_z$ (however, we are limiting ourselves to
$E_z\ll E_F$; the case of strong
fields where the electron system is close to full polarization will be
addressed elsewhere).

{\em Method} -- The expression for the leading interaction correction 
to conductivity 
can be found either by means of the diagrammatic technique \cite{aar}  
or using the quantum kinetic equation \cite{us1}. Both methods are 
completely equivalent and result in the following expression for 
the correction \cite{us1} 

\end{multicols}

\widetext

\begin{mathletters}
\label{zerofield}
\beq
\frac{\delta \sigma_{xx}}{\sigma_{D}}=
\Im \int\limits_{-\infty}^{\infty} \frac{d\w}{\pi}
\frac{\partial}{\partial \w} \left( \w \coth \frac{\w}{2T} \right)
\int\limits_0^{\infty} \frac{qdq}{4\pi} \left[
 \mbox{\bf Tr} \widehat{{\cal{D}}}^R(\w, q) \right]
B_{xx}(\w, q),
\eeq

\noindent
where the retarded interaction propagator $[\widehat{{\cal{D}}}^R(\w,
q)]_{\sigma_1 \sigma_2; \sigma_3 \sigma_4}$ is a matrix in spin space,
and the form-factor $B_{xx}$ is defined as

\beqa
B_{xx}(\omega, q)=
\left\{
\frac{ v_F^2 q^2/\tau^2}{C^3 (C - 1/\tau)^3} 
+ \frac{3 v_F^2 q^2}{2 \tau C^3 (C-1/\tau)^2} 
+ \frac{2[C-(-i\w + 1/\tau)]}{C(C-1/\tau)^2} +
\frac{2C-1/\tau}{C v_F^2 q^2} 
\left( \frac{C-(-i\w + 1/\tau)}{C-1/\tau} \right)^2
\right\}, 
\eeqa
\end{mathletters}

\begin{multicols}{2}

\noindent 
using the notation

\beqa
C(\w, q) = \sqrt{\left(-i\w + 1/\tau\right)^2 + v_F^2 q^2}.
\label{bxx}
\eeqa

\noindent 

In the absence of magnetic field one can choose a basis corresponding
to the singlet (charge) and triplet channels in which the interaction
propagator becomes diagonal $\widehat{{\cal D}}^R = \mbox{diag}({\cal
D}^R_s,{\cal D}^R_t,{\cal D}^R_t,{\cal D}^R_t)$. The interaction
propagator in the singlet channel is taken in the unitary limit (and
hence is independent of the corresponding Fermi-liquid parameter
$F^{\rho}$) and becomes proportional to the inverse of the electronic
polarization operator

\beq
{\cal D}_s^R = -   
\frac{1}{\Pi^R}, 
\label{f-prop} 
\eeq 
\beq
\Pi^R(\w, q) = \nu \left(1 - \frac{-i\w}{C(\w, q) - 1/\tau} \right).
\eeq 

\noindent 
On the contrary, the triplet channel propagator depends on the Fermi-liquid 
constant $F_0^{\sigma}$ 
 
\begin{eqnarray} 
{\cal D}_t^R = - 
\frac{F^\sigma_0}{\nu + F^\sigma_0 \Pi^R} =
- \frac{1}{\nu} 
\frac{C-1/\tau}{i\w + \frac{F_0^\sigma+1}{F_0^\sigma} (C-1/\tau)},
\label{trip-prop} 
\end{eqnarray} 
 
\noindent 
and describes spin-exchange coupling. For details of the derivation of 
the propagators and Eqs.~(\ref{zerofield}) we refer the reader to 
Ref.~\onlinecite{us1}.  
 
Using the explicit form of propagators (\ref{f-prop}) and 
(\ref{trip-prop}) we evaluate the integral Eq.~(\ref{zerofield}) and 
find \cite{us1} the temperature dependent correction to conductivity 
in the absence of external magnetic field: 
 
\begin{mathletters} \label{6}
\begin{equation} 
\sigma = \sigma_D + \delta\sigma_C + \delta\sigma_T. 
\label{r0} 
\end{equation} 
 
\noindent 
Here the charge (singlet) channel contribution is given by 
 
\begin{equation} 
\delta\sigma_C = 
\frac{e^2}{\pi\hbar} \frac{T\tau}{\hbar}  
\left[ 1 -\frac{3}{8}f(T\tau)\right] 
 -\frac{e^2}{2\pi^2\hbar}\ln\left(\frac{E_F}{T}\right), 
\label{fc} 
\end{equation} 
 
\noindent 
and the triplet channel correction is 
 
\begin{eqnarray} 
\delta\sigma_T = && 
\frac{3F_0^\sigma}{(1+F_0^\sigma)} 
\frac{e^2}{\pi\hbar}\frac{T\tau}{\hbar}  
\left[ 1 -\frac{3}{8}t(T\tau; F_0^\sigma)\right] 
\nonumber\\ 
&& 
\nonumber\\ 
&& 
-3\left(1-\frac{1}{F_0^\sigma} 
\ln(1+F_0^\sigma)\right) 
\frac{e^2}{2\pi^2\hbar}\ln\left(\frac{E_F}{T}\right). 
\label{tc} 
\end{eqnarray} 
\label{s} 
\end{mathletters} 
 
\noindent 
The factor of $3$ in the triplet channel correction Eq.~(\ref{tc}) is
due to the fact that all three spin components of the triplet state
contribute equally. The function $f(T\tau)$ smoothly decays from unity
to zero and the function $t(T\tau; F_0^\sigma)$ is non-monotonous only
in the narrow region $-0.25 > F_0^\sigma > -0.5$ where it has a
maximum at $T\tau = 1/(1+F_0^\sigma)$. For numerical reasons both
$f(T\tau)$ and $t(T\tau; F_0^\sigma)$ change the result only by a few
percent and therefore their explicit form (given in
Ref.~\onlinecite{us1}) is inessential for the present discussion.
 
The correction (\ref{s}) is written in the approximation of constant
(i.e. momentum-independent) $F_0^\sigma$.  For the system close to the
Stoner instability such as $1/(\epsilon_F\tau) \ll (1+F_0^\sigma) \ll
1$, this limits the applicability of Eq.~(\ref{6}) by temperatures
smaller than $T^* = \epsilon_F(1+F_0^\sigma)^2$, see Ref.~\onlinecite{us1}.

In parallel magnetic field electrons acquire additional Zeeman energy
$E_z = g \mu_B H$, which is proportional to the magnitude $H$ of the
field, the Bohr magneton $\mu_B$, and the electron $g$-factor.
Consequently, the exact Green's functions of non-interacting electrons
now depend on magnetic field. They are related to the Green's
functions in the absence of the field as
 
\[ 
{G}^{R,A}(\epsilon) 
\to  
{G}^{R,A}\left(\epsilon-\frac{1}{2}E_z\hat{\sigma}_z\right), 
\] 
 
\noindent 
where $\hat{\sigma}_z$ is the Pauli matrix in the spin space, and we
chose the $z$-axis along the direction of the magnetic
field. Repeating all the considerations of Ref.~\onlinecite{us1}, one
finds that two-particle propagators (that depend on the difference of
the electron energies) are also modified by the field. 
This
modification depends on the spin state of the two particles
\cite{aar}.
Consider first a system of non-interacting electrons. Identification
of the singlet and triplet channels corresponds to the choice of a
basis in spin space, namely using the states with the total spin $L$
and its $z$-component $L_z$.  The singlet channel is the state with
$L=0$ and it is unaffected by the magnetic field, as is the $L_z=0$
component of the triplet. For the remaining two components the Zeeman
splitting results in the shift of the frequency $\omega$ in all
diffusons by $L_z E_z$.
 
In the presence of electron-electron interaction one takes into 
account the external magnetic field mostly in the the same manner. The 
only difference is that the $g$ factor is renormalized by the 
spin-exchange interaction similarly to the Pauli susceptibility 
Eq.~(\ref{pauli}). Consequently, the Zeeman splitting is also 
renormalized:

\begin{eqnarray*}
E_z^*=\frac{g\mu_B H }{1+F^\sigma_0}.
\end{eqnarray*}

\noindent 
Thus the conductivity correction (\ref{zerofield}) is modified as: 
  
\begin{mathletters}  
\begin{eqnarray} 
&&\frac{\delta\sigma_{xx} (H)}{\sigma_D} 
= \Im \int\limits_{-\infty}^{\infty}  
\frac{d\omega}{\pi} \frac{\partial}{\partial\omega} 
\left(\omega\coth\frac{\omega}{2T}\right) \int \frac{qdq}{4\pi} 
\nonumber\\  
&& 
\nonumber\\ 
&&\quad\quad\quad\quad 
\times \Bigg\{  
\Big[{\cal D}_s^R(\omega, q)+{\cal D}_t^R(\omega, q)\Big] B_{xx}(\omega,q) 
\label{l0} 
\\ 
&& 
\nonumber\\ 
&&\quad\quad 
+\sum\limits_{L_z=\pm 1} 
{\cal D}_t^R(L_zE_z^*;\omega, q) B_{xx}(\omega + L_zE_z^*,q)  
\Bigg\}; 
\label{l1} 
\end{eqnarray} 
\label{int} 
\end{mathletters} 
 
\noindent 
where the form-factor $B_{xx}(\omega, q)$ is given by Eq.~(\ref{zerofield}), 
propagators in the $L_z=0$ channels expression~(\ref{l0}) are still 
given by Eqs.~(\ref{f-prop}) and (\ref{trip-prop}), while the 
propagators in expression~(\ref{l1}) are modified by the Zeeman energy as 
follows: 
 
\begin{eqnarray} 
{\cal D}_t^R(L_zE_z^*;\omega, q) = - 
\frac{F^\sigma_0}{\nu + F^\sigma_0 \Pi^R(L_zE_z^*;\omega, q)},
\label{l-prop} 
\end{eqnarray} 
\begin{eqnarray} 
&&\Pi^R(L_z E_z^*;\omega, q)=\nu\left[1 -  
\frac{-i\omega}{C(\w+L_z E_z^*,q)
-{1}/{\tau}}\right]. 
\label{l-pol} 
\end{eqnarray} 
 
\noindent 
Note that the numerator of the polarization operator Eq.~(\ref{l-pol}) 
is not changed by the Zeeman energy. As a result, the pole of the 
propagator Eq.~(\ref{l-prop}) at $q=0$ depends only on the bare 
Zeeman energy $E_z$ with the bare electron 
$g$-factor. This is a manifestation of the Larmor theorem: the 
frequency of a homogeneous collective mode (which is the meaning of 
the pole at $q=0$) can not be renormalized by electron-electron 
interaction. 
 
{\em Results} -- Given the expression 
for the correction Eq.~(\ref{int}) and the 
explicit expression for the triplet propagator in the presence of the 
Zeeman field Eq.~(\ref{l-prop}), further calculation consists 
of evaluating the integral in Eq.~(\ref{int}). The integral is similar to
its zero-field counterpart (see Ref.~\onlinecite{us1}). The resulting
magneto-conductivity can be written as
 
\begin{eqnarray} 
\sigma(H,T)&-&\sigma(0,T) =  
\frac{e^2}{\pi\hbar} 
\left[ 
\frac{2F_0^\sigma}{(1+F_0^\sigma)}\frac{T\tau}{\hbar} 
{\; }K_{b}\left(\frac{E_z}{2T}, F_0^\sigma\right)\right. 
\nonumber
\\ 
& & 
\nonumber\\ 
&+& \left. K_{d}\left(\frac{E_z}{2\pi T}, F_0^\sigma\right)
+ m(E_z\tau, T\tau; F_0^\sigma) \right]. 
\label{mr}
\end{eqnarray} 
 
\noindent 
In the ballistic limit $T\tau\gg 1$ the dominating contribution is
given by the first term in Eq.~(\ref{mr}), where the dimensionless
function $K_{b}(x, F_0^\sigma)$ contains two contributions:

\begin{mathletters} 
\begin{eqnarray} 
K_{b}(x, F_0^\sigma) = K_{1}(x) + K_{2}(x, F_0^\sigma),
\label{b1} 
\end{eqnarray} 

\noindent
where 

\begin{eqnarray} 
K_{1}(x) = x\coth x -1,
\label{b2} 
\end{eqnarray} 

\noindent
and

\begin{eqnarray} 
K_{2}(x, F_0^\sigma)&& = \frac{1+F_0^\sigma}{2F_0^\sigma}
\int\limits_x^{
{x}/{(1+F_0^\sigma)}} 
dy \frac{\partial}{\partial y}\Big(y\coth y\Big)
\label{b3}
\\
&&
\nonumber\\
&&
\times\left(y-\frac{x}{1+F_0^\sigma}\right)
\left[\frac{1}{y}+\frac{2F_0^\sigma}{(1+2F_0^\sigma)y - x}\right].
\nonumber
\end{eqnarray} 
\label{bal}
\end{mathletters}

If the magnetic field is strong, $x\gg 1$, the expression Eq.~(\ref{bal})
simplifies to

\beq
K_{b}(x\gg 1,F_0^\sigma) = g(F_0^\sigma) x - 1 
+ {\cal O} \left( \frac{1}{x} \right), 
\eeq

\noindent
where the dimensionless function $g(z)$, not to be confused with the
Lande $g$-factor, is

\beqa
g(&&z) =  
\frac{1}{2z}\ln(1+z)
+ \frac{1}{2(1+2z)} 
+ \frac{z\ln2(1+z)}{(1+2z)^2}.
\nonumber
\end{eqnarray}

\noindent
For the smallest magnetic field, $x\ll 1+F_0^\sigma$, we have

\begin{eqnarray}
K_{b}(x\ll 1+F_0^\sigma, F_0^\sigma)\approx \frac{x^2}{3} f(F_0^\sigma),
\end{eqnarray}

\noindent
where

\begin{eqnarray*}
f(&&z) = 1 - 
\\
&&
\nonumber\\
&& 
-\frac{z}{1+z} \left[ \frac{1}{2} + \frac{1}{1+2z}
-\frac{2}{(1+2z)^2}+\frac{2\ln2(1+z)}{(1+2z)^3}\right].
\end{eqnarray*}

\noindent 
The diffusive limit $T\tau \ll 1$ is characterized by the function 
 
\end{multicols} 
\widetext 
 
\begin{eqnarray} 
&&K_{d}(h,F_0^\sigma) = - \frac{1}{4\pi F_0^\sigma} 
\sum\limits_{L_z=\pm 1}{\rm Re}{\; } 
\int\limits_{-\infty}^{\infty} \frac{d x}{x^2}  
\left[\frac{\partial}{\partial x}\Big(x\coth \pi x\Big) \right] 
\Big(x - L_zh\Big) \ln \frac{x - L_zh}{x - L_zh/(1+F_0^\sigma)} 
\label{diff} 
\\ 
&& 
\nonumber\\ 
&& 
=-\frac{1}{2 \pi F_0^\sigma} \sum\limits_{n=1}^{\infty} 
\left\{ \frac{1}{n}\left(\ln\frac{n^2+h^2}{n^2 
+\frac{h^2}{(1+F_0^\sigma)^2}}  
\right) 
-\frac{4h}{n^2}\left(\arctan\frac{h}{n} - \arctan\frac{h}{n(1+F_0^\sigma)} 
\right)\right\} 
- \frac{1}{\pi}\Big[ {\cal C} + {\rm Re \ }  
\psi\left(1-\frac{ih}{1+F_0^\sigma}\right) \Big], 
\nonumber 
\end{eqnarray} 
 
\begin{multicols}{2} 
\noindent 
where ${\cal C}=0.577\dots$ is Euler's constant, and $\psi(x)$ is the 
di-gamma function. For weak interaction ($F_0^\sigma\ll 1$)
Eq.~(\ref{diff}) reproduces the result of Ref.~\onlinecite{lrp}. 
At the smallest magnetic field $h \ll 1+F_0^\sigma$ 
Eq.~(\ref{diff}) reduces to 
 
\begin{eqnarray} 
K_{d}(h) \approx \frac{3F_0^\sigma \zeta(3)}{2\pi(1+F_0^\sigma)^2} 
h^2, 
\label{smh} 
\end{eqnarray} 
 
\noindent 
where $\zeta(x)$ is the Riemann zeta-function,  
$\zeta(3)=1.202\dots$. 
In the opposite limit $h \gg 1$ we have 
 
\begin{eqnarray} 
&& K_{d}(h) \approx 
\frac{1}{\pi} 
\left\{ 
1 - \frac{1}{F_0^\sigma} \ln (1 + F_0^\sigma) 
\right\}
\ln \frac{h}{1+F_0^\sigma} 
 \nonumber \\
&& \quad \quad \quad - \frac{1}{2\pi F_0^\sigma} 
\ln^2 \frac{1}{1+F_0^\sigma}. 
\label{lrh} 
\end{eqnarray} 
 
\noindent 
Finally, for intermediate values $1+F_0^\sigma \ll h \ll 1$ we obtain

\begin{equation} 
K_{d}(h) \approx - \frac{1}{2\pi} \ln^2 \frac{h}{1+F_0^\sigma} +  
{\cal O} \left( \ln \frac{h}{1+F_0^\sigma} \right). 
\end{equation} 
 
The cross-over between the ballistic and the diffusive regimes is
described by the dimensionless function $m(E_z\tau, T\tau;
F_0^\sigma)$.  In the absence of the field $m(0, T\tau; F_0^\sigma) =
0$.  Similarly to the function $t(T\tau; F_0^\sigma)$ in
Eq.~(\ref{tc}), this function appears to be numerically small and does
not modify the sum of the two limiting expressions Eqs.~(\ref{diff})
and (\ref{bal}) by more than one per cent.

{  
\narrowtext  
\begin{figure}[ht]  
\epsfxsize=7 cm  
\centerline{\epsfbox{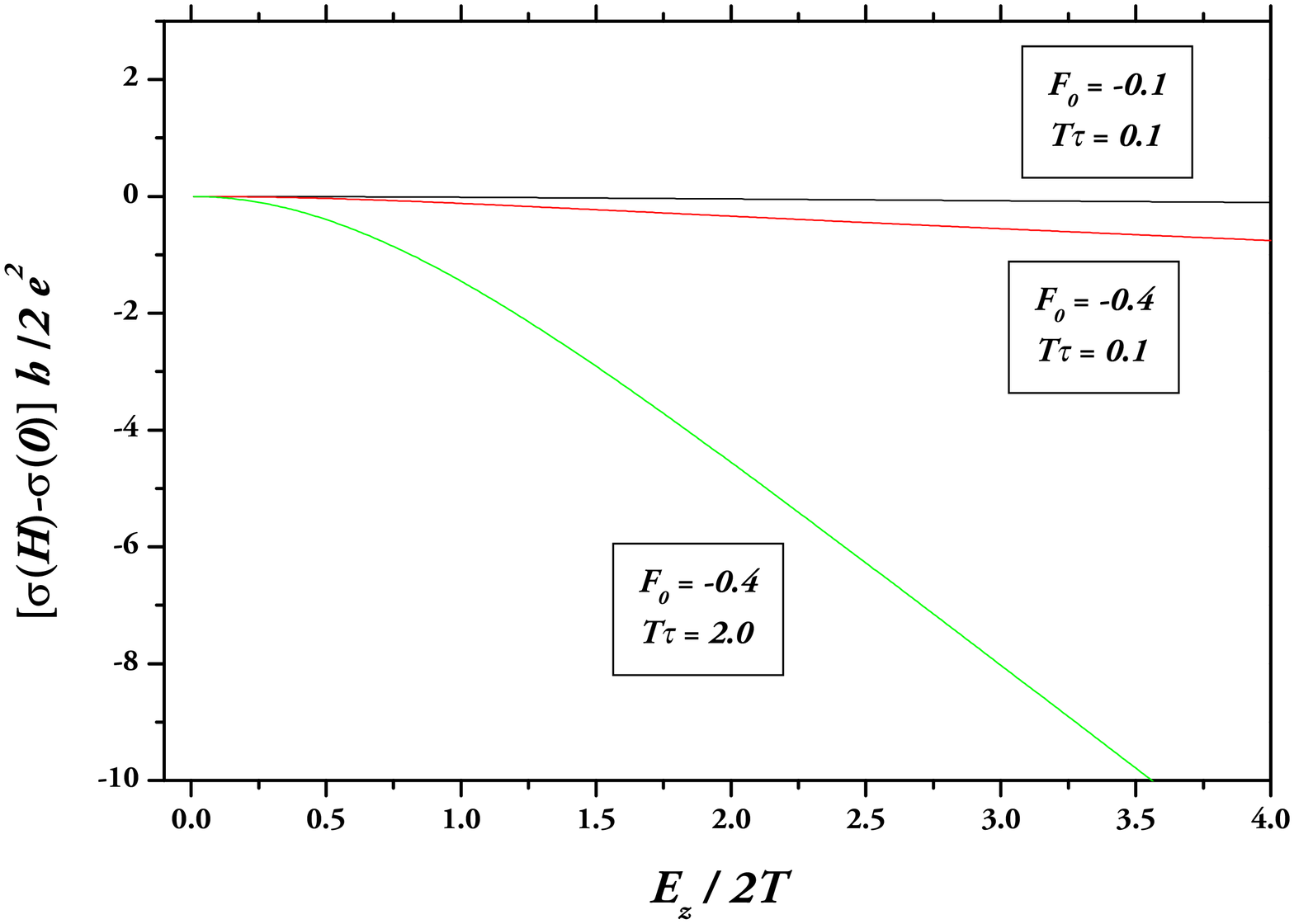}} 
\caption{Magneto-conductivity in parallel field for different values of
$F_0^\sigma$.}  
\label{1} 
\end{figure} 
} 

{ 
\narrowtext
\begin{figure}[ht]  
\epsfxsize=7 cm  
\centerline{\epsfbox{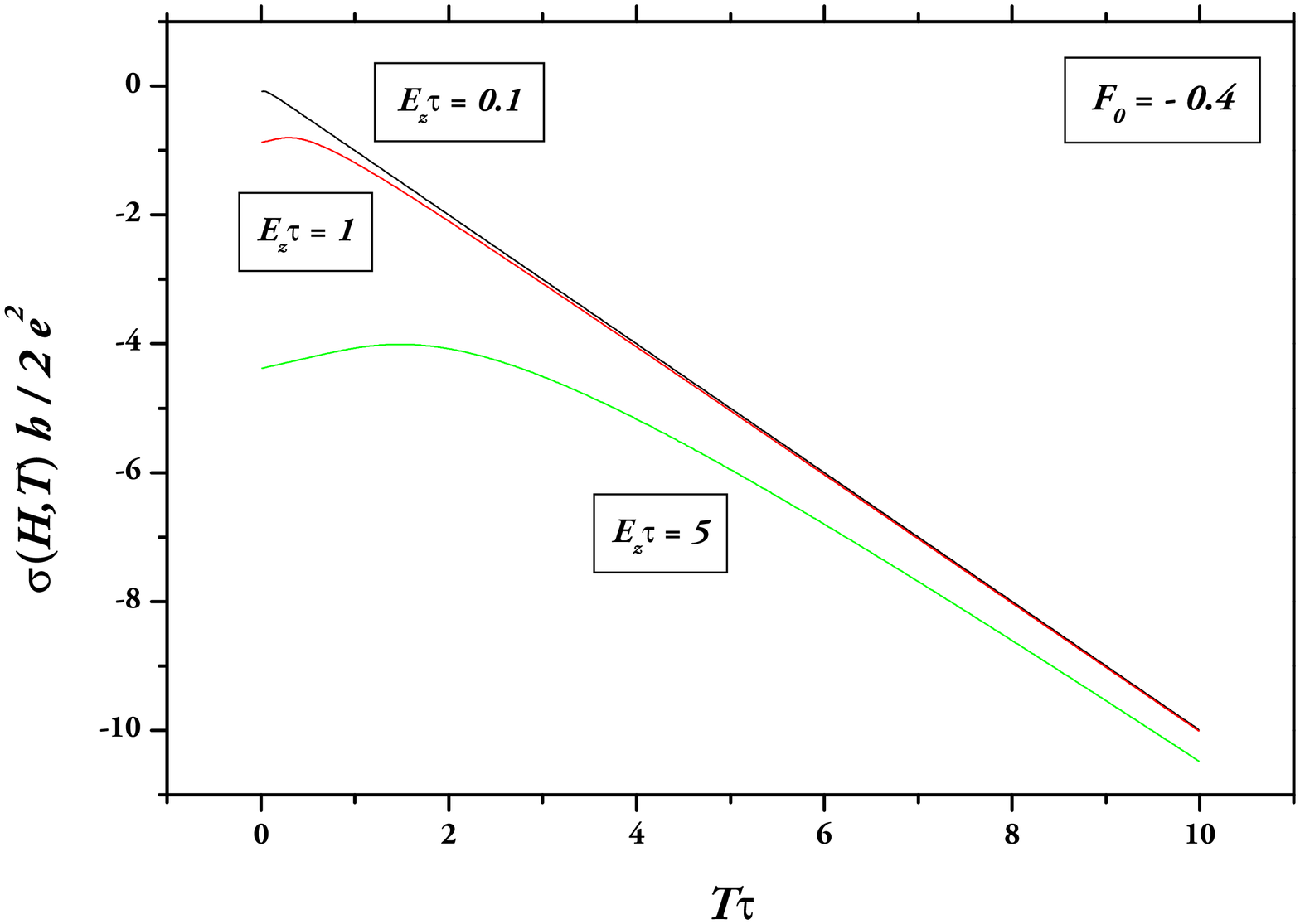}} 
\caption{Temperature dependence of the conductivity corrections in the
presence of the parallel field. }  
\label{2} 
\end{figure} 
} 

{\em Discussion} -- The resulting temperature dependence 
of the conductivity correction
Eq.(\ref{int}) is summarized in Figs.~\ref{2} and \ref{3}. In the 
ballistic regime $\delta\sigma\propto T$. Remarkably, the value and 
{\it the sign} of the slope depends on the field. At zero field 
\cite{us1} the correction is given by all four (the singlet and three 
components of the triplet) spin  channels so that 
$\partial \sigma/\partial T \propto 1 + 3F^{\sigma}_0/(1+F^{\sigma}_0)$. 
For stronger fields $E_z > T$ the $L_z=\pm 1$ channels are gapped and 
we are left with one singlet and one triplet channel
$\partial \sigma/\partial T \propto 1 + F^{\sigma}_0/(1+F^{\sigma}_0)$.
The cross-over is described by Eqs.(\ref{bal}) and shown on Fig.~\ref{3}.
This picture is valid up to fields of order $(1+F_0^\sigma)^2 E_F$. 
At the strongest fields $E_z^* > E_F$ when the system is fully 
polarized  the spin does not play a role any more and one retrieves the 
universal singlet channel result [see Eq.~(\ref{fc})]

\begin{equation}
\frac{\partial\sigma}{\partial T} = \frac{e^2\tau}{\pi \hbar^2};
\quad\quad
\frac{T\tau}{\hbar}\gtrsim 0.1;
\quad
E_z^* > E_F.
\end{equation} 

\noindent
This conclusion is in 
agreement with recently reported measurements of magneto-resistance in
{\it GaAs} heterostructures \cite{sav}.

{ 
\narrowtext
\begin{figure}[ht]  
\epsfxsize=7 cm  
\centerline{\epsfbox{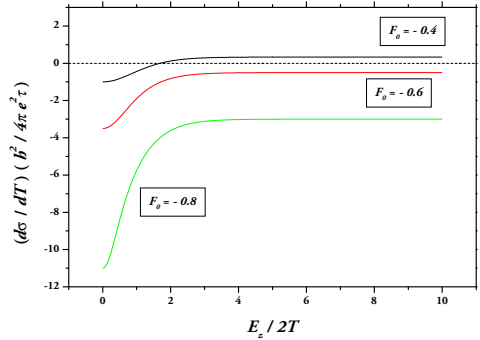}} 
\caption{Slope of the temperature dependence of the conductivity correction
(in the ballistic limit) as a function of the parallel magnetic field}  
\label{3} 
\end{figure} 
}

{\em Acknowledgments} -- We are grateful to 
B.L. Altshuler, V. Falko, S.V. Kravchenko,
and A.K. Savchenko for stimulating discussions.
One of us (I.A.) was supported by the Packard foundation. Work at Lancaster
University was partially funded by EPSRC-GR/R01767.

\end{multicols} 
 
\end{document}